\documentclass[conference]{IEEEtran}
\IEEEoverridecommandlockouts
\usepackage{cite}
\usepackage{amsmath,amssymb,amsfonts}
\usepackage{algorithmic}
\usepackage{graphicx}
\usepackage{textcomp}
\usepackage{xcolor}
\def\BibTeX{{\rm B\kern-.05em{\sc i\kern-.025em b}\kern-.08em
    T\kern-.1667em\lower.7ex\hbox{E}\kern-.125emX}}

\usepackage{amsmath,amssymb,amsfonts}
\usepackage{algorithm}
\usepackage{algorithmic}
\usepackage{tabularx}
\usepackage{graphicx}
\usepackage{textcomp}
\usepackage{xcolor}
\usepackage{afterpage}

\begin{document}
\setlength{\columnsep}{0.2in} 

\title{Serv-Drishti: An Interactive Serverless Function Request Simulation Engine and Visualiser}

\author{\IEEEauthorblockN{Siddharth Agarwal, Maria A. Rodriguez, and Rajkumar Buyya}
\IEEEauthorblockA{\textit{Cloud Computing and Distributed Systems (CLOUDS) Laboratory} \\
\textit{School of Computing and Information Systems}\\
\textit{The University of Melbourne, Australia}\\
siddhartha@student.unimelb.edu.au, \{maria.read, rbuyya\}@unimelb.edu.au}
}

\maketitle

\begin{abstract}
The rapid adoption of serverless computing necessitates a deeper understanding of its underlying operational mechanics, particularly concerning request routing, cold starts, function scaling, and resource management.  This paper presents Serv-Drishti, an interactive, open-source simulation tool designed to demystify these complex behaviours. Serv-Drishti simulates and visualises the journey of a request through a representative serverless platform, from the API Gateway and intelligent Request Dispatcher to dynamic Function Instances on resource-constrained Compute Nodes. Unlike simple simulators, Serv-Drishti provides a robust framework for comparative analysis. It features configurable platform parameters, multiple request routing and function placement strategies, and a comprehensive failure simulation module. This allows users to not only observe but also rigorously analyse system responses under various loads and fault conditions. The tool generates real-time performance graphs and provides detailed data exports, establishing it as a valuable resource for research, education, and the design analysis of serverless architectures.
\end{abstract}


\section{Introduction}
Serverless computing has emerged as a transformative paradigm, abstracting away the complexities of infrastructure management and enabling developers to focus solely on their application logic. This model, often referred to as Function-as-a-Service (FaaS), has seen widespread adoption due to its inherent benefits of elastic scalability, cost-effectiveness through a pay-as-you-go model, and reduced operational overhead \cite{DreScale}. Major cloud providers such as Amazon Web Services (AWS) \cite{awsLambda}, Google Cloud \cite{runFunctions}, and Microsoft Azure \cite{shahradArchitecture2019} have made serverless computing a fundamental part of modern cloud-native architecture. However, the very abstraction that makes serverless attractive also introduces a significant challenge: the lack of transparency \cite{DreScale}. The internal mechanisms governing request dispatching, resource provisioning, and dynamic scaling are often treated as a \textit{black-box}. This leaves developers to deal with the unexpected delays of a cold start \cite{ICCIDA_Siddharth}, the complexities of elastic scaling \cite{DreScale}\cite{shahradArchitecture2019}, and the nuances of request routing logic \cite{CloudsimSC}. Understanding these behind-the-scenes processes is crucial for optimising serverless application performance, predicting behaviour under load, and designing resilient, cost-effective systems.

Existing approaches \cite{PerformanceModellingSC}\cite{PerformanceModellingMetricBased} predominantly focus on performance monitoring, modelling, and deployment, leaving a significant gap for demonstrative and educational instruments that visually explain the dynamic lifecycle of a serverless request. To address this gap, we introduce  \textbf{\textit{Serv-Drishti}}, an interactive, end-to-end serverless workflow simulation engine and visualiser. By providing a lucid, hands-on experience, our tool demystifies serverless operations for students, researchers, and practitioners. This paper details the architecture, simulation logic, and features of Serv-Drishti, including its advanced failure simulation, performance analysis, and data export modules, demonstrating its value as a powerful platform for understanding and analysing serverless architectures.

\section{Related Work}
The rapid evolution of serverless computing has driven considerable research into modelling, simulation, and performance analysis of FaaS platforms to facilitate deeper understanding. Therefore, simulators are crucial for comprehending scheduling behaviour and resource management decisions in FaaS environments without incurring cloud costs. 

\textbf{CloudSimSC} \cite{CloudsimSC} builds on the CloudSim \cite{Cloudsim7G} toolkit for serverless environments, modelling detailed resource management and scheduling policies. It delivers robust trace-driven experimentation for evaluating concurrent request handling and scaling strategies, but operates in a non-visual manner. Its outputs are logs and post-processed metrics, making it less suitable for immediate interactive demonstration or pedagogical visualisation of request flows.

\textbf{DSLab FaaS} \cite{DSLabFaaS} provides modular, trace-driven FaaS simulation focusing on reproducibility and extensibility. It supports custom plugin components (e.g., schedulers, auto-scalers) and has demonstrated efficient simulations for complex workloads. DSLab FaaS excels at rigorous, repeatable resource management policy evaluation but, like CloudSimSC, does not offer real-time visualisations. Additionally, the workload traces are identified as the source of simulation data and may not support generation of simulated data. 

\textbf{faas-sim} \cite{faassim} is a discrete-event, trace-driven framework built on SimPy, unique for integrating network latency modelling via Ether. faas-sim allows researchers to assess scheduling and autoscaling strategies, especially in distributed or edge environments, but its use of trace analysis and lack of animated visualisation limits accessibility for comparative and pedagogical study.

\textbf{OpenDC} \cite{OpenDCFaaS} is a notable simulation platform for modelling emerging cloud datacenter technologies, including serverless workloads. It provides a valuable environment for exploring different resource allocation and scheduling policies. While OpenDC offers an interactive web interface for model exploration, its visualisations are primarily focused on aggregate metrics and static topology maps rather than dynamically animated request flow. This key distinction is crucial for understanding the impact of scheduling and routing decisions on individual requests, which is central to analysing system bottlenecks and cold starts.

\textbf{ServlessSimPro} \cite{ServerlessSimPro}, a comprehensive simulation platform,  addresses many of the shortcomings of prior simulators. 
It distinguishes itself by offering a wide range of scheduling strategies, including container migration and reuse, and provides a comprehensive set of performance metrics, notably including the first-ever monitoring of energy consumption. 
While ServlessSimPro provides a robust, code-based simulation environment, its core utility remains within a backend framework. The reliance on code execution to define experiments and visualise results presents a steep learning curve and limits accessibility for a broad audience of researchers, students, and practitioners. The purely code-based interface also hinders an intuitive, real-time understanding of the temporal dynamics of a serverless request flow.

\begin{figure}
    \centering
    \includegraphics[width=1\linewidth]{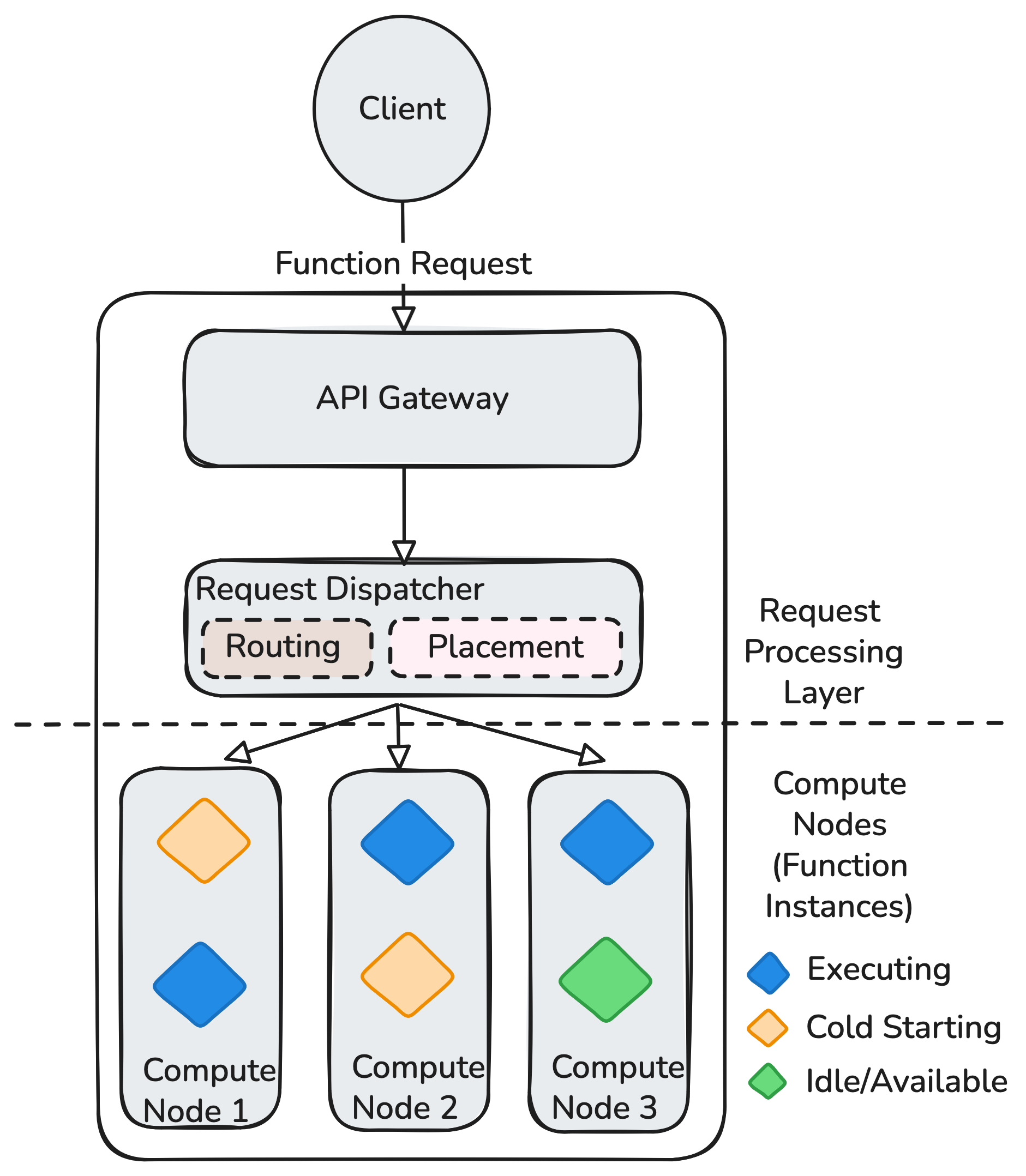}
    \caption{Request Flow across Serv-Drishti Components}
    \label{fig:components}
\end{figure}

\section{System Architecture and Simulation Model}
The visualiser's architecture abstracts the core components of a serverless platform into a simplified, yet functionally representative, model. The simulation is built around a discrete-event, request-driven model, where each incoming request is treated as a unique entity traversing the system and interacting with various components, triggering state changes and resource consumption events. This approach allows for detailed tracking of individual request latencies and resource usage across the system.

The core of the Serv-Drishti platform is its simulation engine, which models the end-to-end request flow through a serverless environment by abstracting key operational components into a layered architecture. Each component is responsible for a distinct phase of the request lifecycle, accurately reflecting the complexities of real-world FaaS platforms.

The \textbf{API Gateway} serves as the initial entry point for all incoming requests, which originate from user actions or automated rates. It forwards requests to the \textbf{Request Dispatcher}, the central intelligence unit of the simulation. The dispatcher manages an internal FIFO Request Queue for buffering requests that cannot be immediately routed to an available function. It applies configurable routing policies, such as \textit{Warm Priority}, \textit{Round Robin}, and \textit{Least Connections}, to select the most suitable function instance. When new capacity is needed, the dispatcher triggers auto-scaling and uses a Placement Algorithm to provision new functions on an available Compute Node.

\textbf{Compute Nodes} represent the underlying infrastructure that hosts one or more \textbf{Function Instances}. They have a configurable, finite capacity of CPU and Memory that is pooled and shared among all the functions they host. Function Instances are the isolated execution environments for the serverless code, capable of processing requests concurrently up to a configurable limit. Each instance consumes a fixed amount of resources from its host node and transitions between states like cold-starting, busy, and active based on system events. This layered architecture allows Serv-Drishti to model the complete request flow and demonstrate the interplay between logical components and physical resources.

\section{Core Simulation Logic and Features}

The core of our platform's simulation logic is the interplay between Request Routing Strategies and Function Placement Algorithms. Request routing governs how incoming requests are dispatched to a function, while function placement determines where a function instance is provisioned on the available virtual nodes. Together, these two mechanisms define how workloads are managed, impacting performance, cost, and resource utilisation.

\subsection{Request Routing Strategies}

The request dispatcher module implements multiple configurable routing strategies to enable users to analyse and compare their effects. The Warm Priority strategy is designed to minimise latency by prioritising the reuse of active ("warm") function instances. A new request is immediately routed to a warm instance if one is available. If no warm instances exist, the request is queued until an instance becomes ready. This approach is common in real-world FaaS platforms to reduce the overhead of cold starts. The Round Robin algorithm, in contrast, is a simple, stateless method that sequentially distributes requests among all currently available instances. While it evenly spreads the load, it does not prioritise warm instances, which can lead to more frequent cold starts, particularly during sudden bursts of traffic. A more intelligent, state-aware algorithm is Least Connections, which routes a new request to the function instance with the fewest concurrent requests. This dynamic load-balancing approach prevents any single instance from becoming a bottleneck, aiming to minimise latency by utilising the least burdened resources.

\subsection{Function Placement Algorithms}
\begin{figure}
    \centering
    \includegraphics[width=1\linewidth]{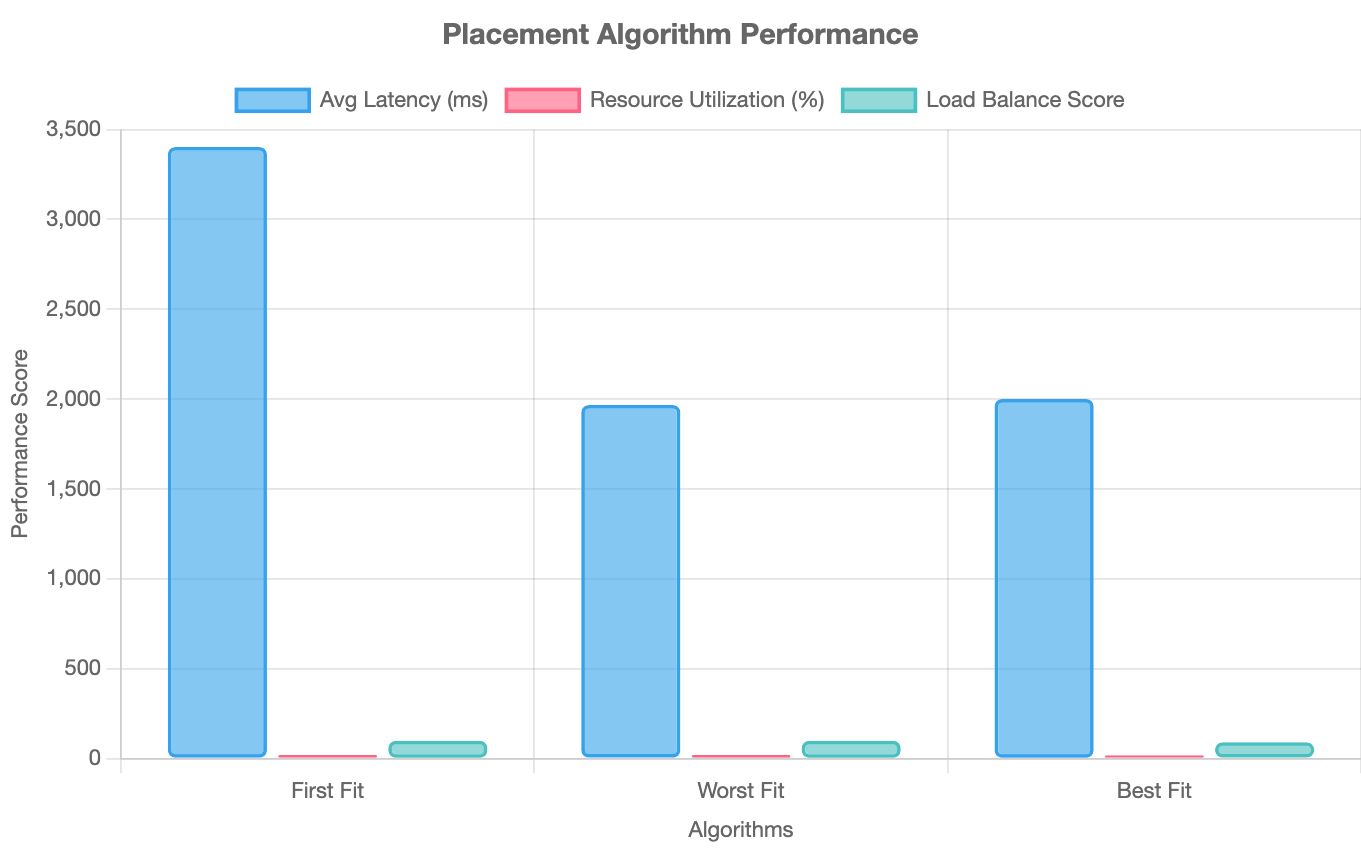}
    \caption{Different Placement Algorithm Performance Comparison}
    \label{fig:placement}
\end{figure}

When a new function instance needs to be created, the system uses a placement algorithm to decide which virtual node will host it. This decision is crucial for optimising resource utilisation, cost, and performance. The First-Fit algorithm places the new instance on the first suitable node found with enough resources. The Best-Fit algorithm selects the node that will have the least remaining capacity after placement, aiming to minimise resource fragmentation. Conversely, the Worst-Fit algorithm places the new instance on the node with the most remaining capacity, leaving room for larger future placements. For balancing the load, the Load-Balanced algorithm places the instance on the node with the lowest average CPU and memory utilisation. The Affinity strategy prefers to place the new instance on a node already hosting a function of the same type to improve resource sharing. Its counterpart, Anti-Affinity, prefers to place the instance on a node that does not host a function of the same type, increasing fault tolerance and isolating workloads. Lastly, the Cost-optimised algorithm is a more complex strategy that seeks to maximise resource utilisation while minimising waste, aiming for the most cost-effective placement decision. A sample comparison of different placement strategies is shown in Fig. \ref{fig:placement}.



\subsection{Function and Compute Node Management}

The lifecycle and state of both function instances and compute nodes are critical elements visually represented in the simulation. A function instance dynamically changes colour to indicate its status. It is \textit{Orange} when in the cold start phase, simulating the time required for initialisation, during which it is unavailable for processing. An instance is \textit{Blue} when it is actively processing requests, up to its configurable concurrency limit, and is \textit{Green} when it is available or warm, ready to receive new requests with minimal latency, Fig. \ref{fig:components}. Compute Nodes represent the underlying physical or virtual machines with a configurable total CPU and memory capacity. A new node is dynamically provisioned by the request dispatcher when existing nodes are at capacity or lack sufficient resources to host new function instances. Each Function Instance consumes a fixed amount of resources, which is deducted from the node's capacity, and the visualiser updates the resource meters in real-time.

\begin{figure}
    \centering
    \includegraphics[width=1\linewidth]{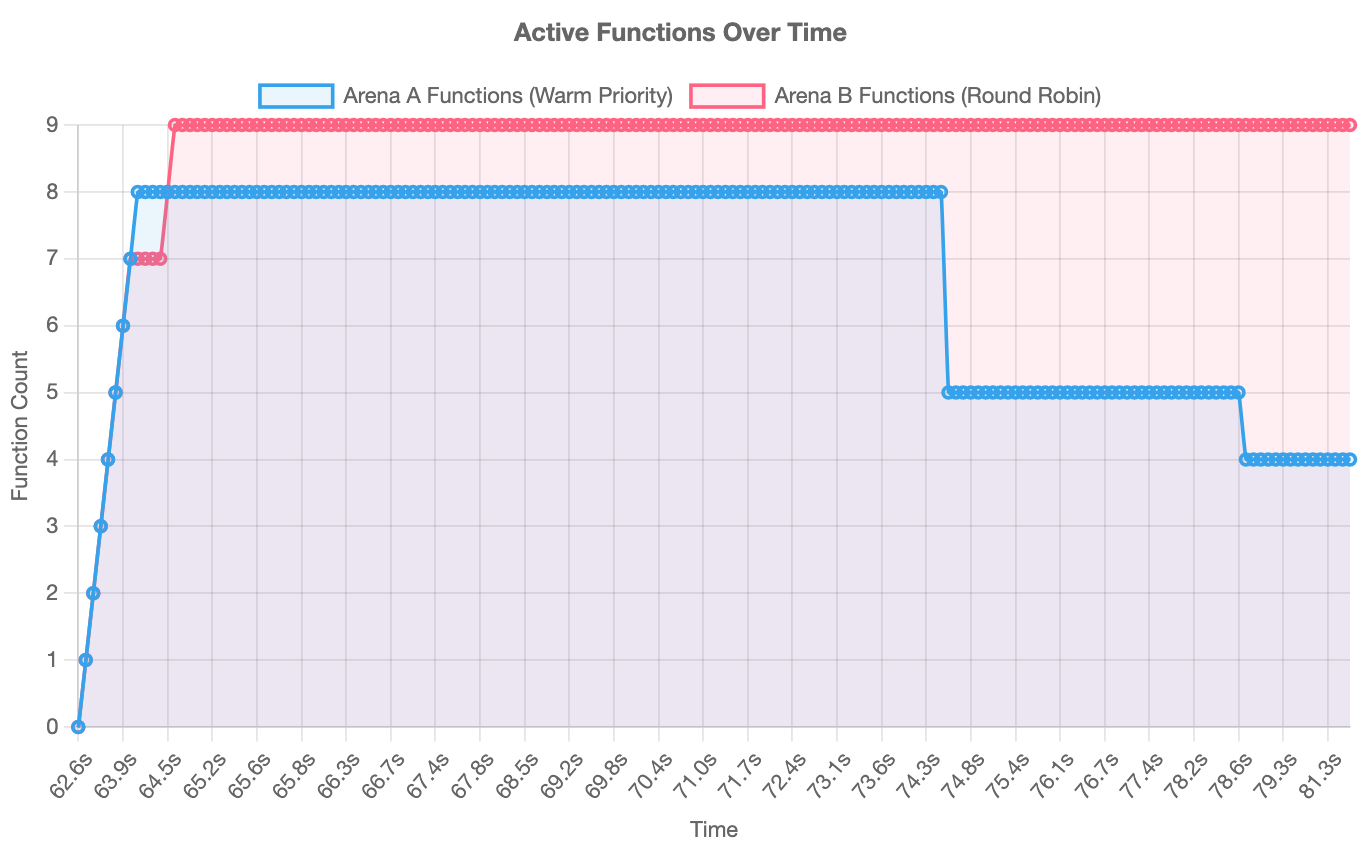}
    \caption{Function Scaling Over Simulation}
    \label{fig:scaling}
\end{figure}

\subsection{Scaling Behaviour}

Serv-Drishti vividly demonstrates both the scale-up (provisioning) and scale-down (de-provisioning) aspects of elasticity, Fig. \ref{fig:scaling}. When demand increases, new function instances are provisioned. If existing compute nodes lack capacity, new nodes are brought online, a process that respects user-defined limits on the maximum number of instances and nodes. To optimise costs, idle function instances and compute nodes are automatically de-provisioned after a configurable Inactivity Timeout. This simulates the pay-as-you-go model by reclaiming idle resources when they are no longer in use.

\subsection{Visualisation and Interaction}

The platform's core strength is its interactive virtualisation, which provides an intuitive understanding of the simulated environment. The request traversal through the system is animated, providing a clear visual of their journey and highlighting potential bottlenecks. The request dispatcher also prominently displays a visual queue, offering immediate feedback on system load and backpressure. A user-friendly, collapsible UI panel allows for real-time adjustments of key simulation parameters, enabling \textit{what-if} scenario analysis and fostering experimental learning in a risk-free environment.

\section{Critical System Considerations}

The design of Serv-Drishti is based on several key considerations aimed at balancing pedagogical value with a representative, yet simplified, simulation of serverless operations. This section delves into these design philosophies and the implications of our approach.

\subsection{Abstraction versus Fidelity}

A fundamental design decision was to create an abstract model rather than a high-fidelity replica of a specific cloud provider's implementation. Real-world FaaS platforms involve immense complexity, including multi-tenant scheduling and intricate networking, which would render the simulator overly complex and difficult for learners to grasp. Instead, Serv-Drishti abstracts away the underlying hardware to highlight the primary logical interactions: request queuing, dispatching, cold starts, concurrent execution, and dynamic scaling. For example, compute nodes abstract the physical infrastructure, while function instances represent the isolated execution environment. This simplification is intentional, as it allows users to focus on fundamental principles of serverless elasticity and resource management, providing a conceptual understanding that is transferable across different FaaS providers. This approach makes the tool highly valuable for educational purposes.

\begin{figure}
    \centering
    \includegraphics[width=1\linewidth]{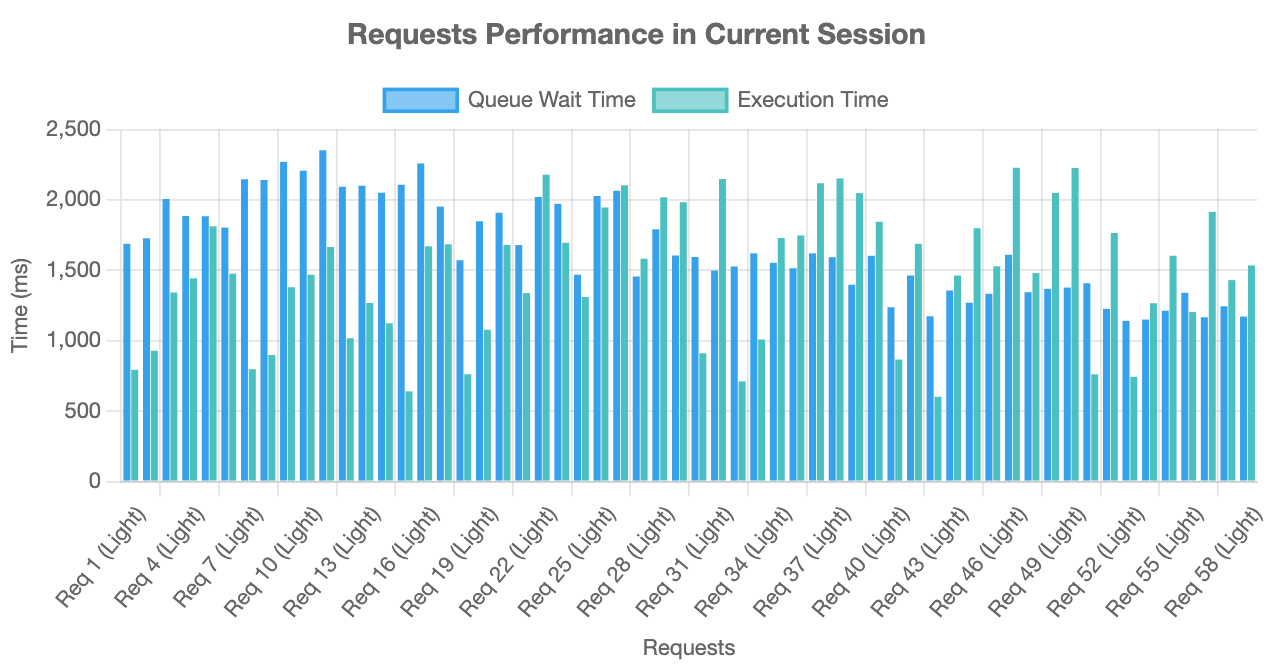}
    \caption{Impact of Request Queue and Cold Start on Function Request Performance }
    \label{fig:latest_performance}
\end{figure}

\subsection{Configurability and Experimental Design}

The extensive configurability of Serv-Drishti is a significant strength for both education and architectural analysis. By allowing users to adjust parameters in real-time, the visualiser becomes a powerful tool for what-if scenario analysis and hypothesis testing. Users can vary the \textit{cold start delay} to understand its impact on latency, especially for different programming languages. The ability to switch between Warm Priority, Round Robin, and Least Connections routing strategies allows for direct comparative analysis of their performance under different load patterns, Fig. \ref{fig:latest_performance}. Furthermore, users can adjust CPU and memory consumption to explore resource contention and the trade-offs between instance size and cost efficiency. Modifying the \textit{max concurrent requests} per function helps in understanding how this parameter affects an instance's utilisation and the system's scaling behaviour. This dynamic experimentation empowers learners and allows architects to rapidly prototype and evaluate design decisions in a risk-free environment.

\subsection{Virtualisation as a Tool for Insight}

The real-time, animated visualisation in Serv-Drishti is a core functional element designed to enhance comprehension and intuition. 
It leverages them by animating request journeys, seeing requests moving through the system, queuing, and changing function states. This provides an immediate and intuitive understanding of the workflow. The instant visual updates of queue length, function states, and node resource meters provide real-time feedback that allows users to directly correlate parameter changes with system behaviour. This feedback loop facilitates active learning and reinforces conceptual understanding. The visual queue and colour-coded instances immediately draw attention to potential bottlenecks, allowing users to quickly identify problematic configurations or load conditions.

\begin{figure}
    \centering
    \includegraphics[width=1\linewidth]{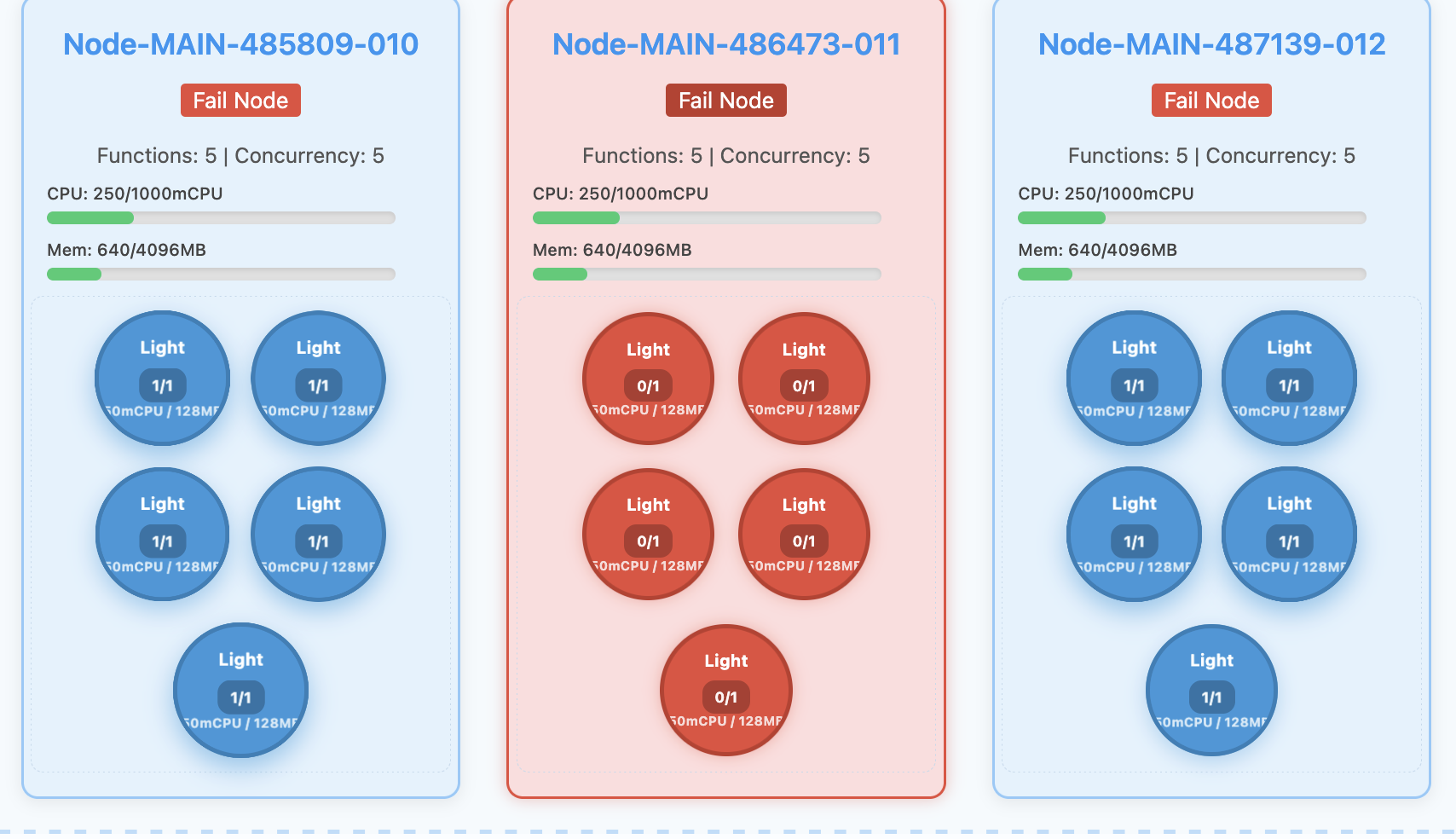}
    \caption{A Failure Simulation through Fail Node}
    \label{fig:fail_node}
\end{figure}

\section{Failure Simulation and Robustness}

Understanding system resilience is paramount in distributed and cloud-native architectures. Serv-Drishti integrates a robust failure simulation module, enabling users to observe and analyse the system's response to various fault conditions, which provides invaluable insights into designing resilient serverless applications and understanding the importance of failure handling mechanisms. The simulation models several key failure scenarios. Each queued request is assigned a configurable Time-to-Live (TTL). If it is not processed within this limit, it is marked as failed and removed from the queue, which simulates real-world client timeouts and demonstrates the impact of unfulfilled requests due to system congestion or delays. Additionally, each Function Instance has a configurable \textit{maximum execution timeout}. If a request exceeds this duration, all requests on that function are marked as failed, and the instance may be terminated, which highlights the importance of setting appropriate timeouts to prevent long-running executions from consuming excessive resources. Finally, the visualiser provides a "Fail Node" button, Fig. \ref{fig:fail_node}, allowing users to manually trigger an immediate infrastructure failure. When a node fails, all hosted functions and in-flight requests on those functions are marked as failed, but the system's robustness is demonstrated as the request dispatcher's logic automatically routes new requests to healthy nodes and provisions new function instances on them, if capacity allows. Failure events are clearly marked visually within the simulation and are accurately reflected in the performance graphs, providing a clear and compelling demonstration of failure propagation and the importance of designing for transient failures.

\section{Performance Analysis and Data Export}

Beyond its visual and interactive capabilities, Serv-Drishti provides quantitative data for a deeper, more rigorous performance analysis, essential for both academic research and practical architectural design. The platform integrates a robust metrics and analytics engine that provides real-time data and comprehensive reports.

\begin{figure}
    \centering
    \includegraphics[width=1\linewidth]{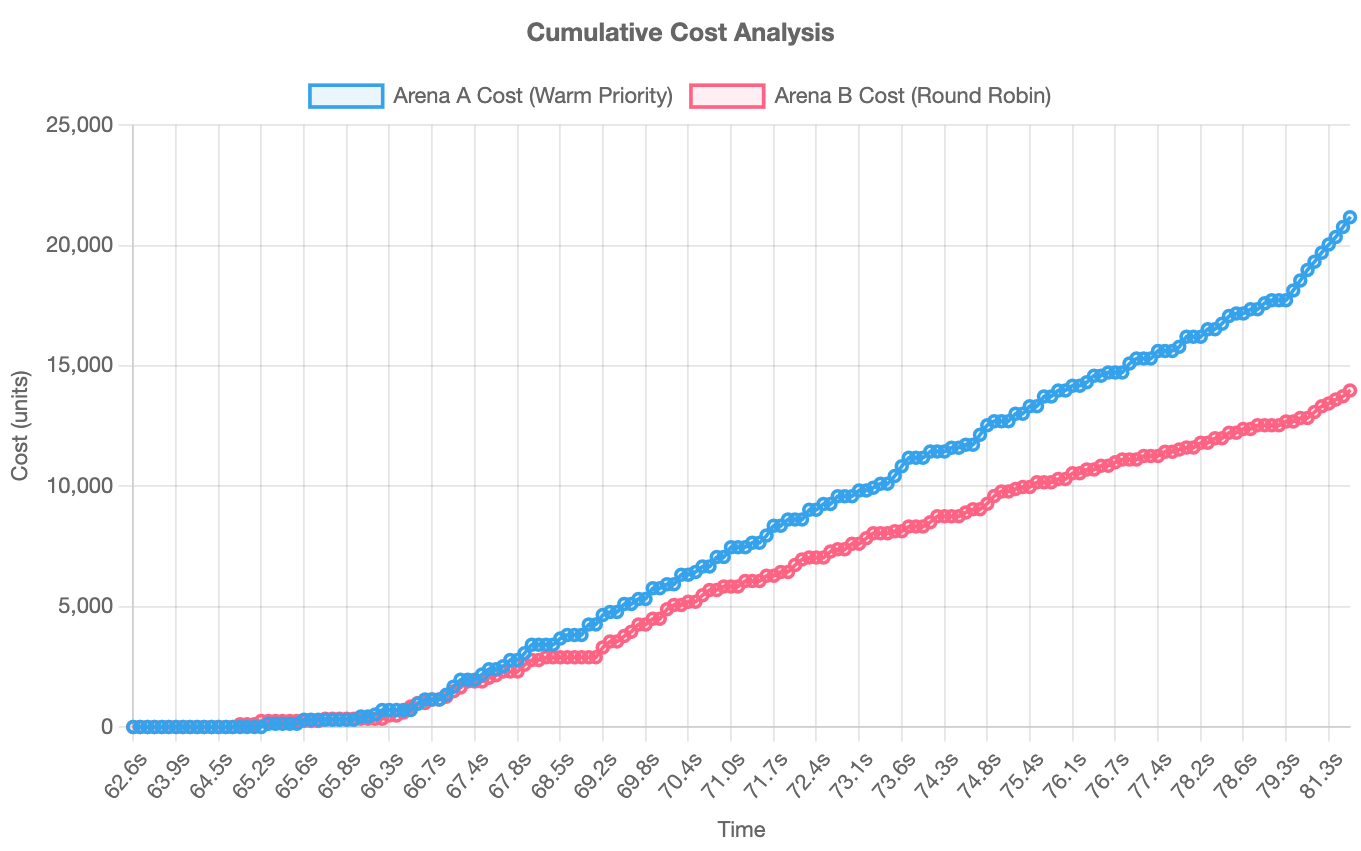}
    \caption{Cumulative Cost Analysis in Serv-Drishti}
    \label{fig:cost}
\end{figure}

\begin{figure}
    \centering
    \includegraphics[width=1\linewidth]{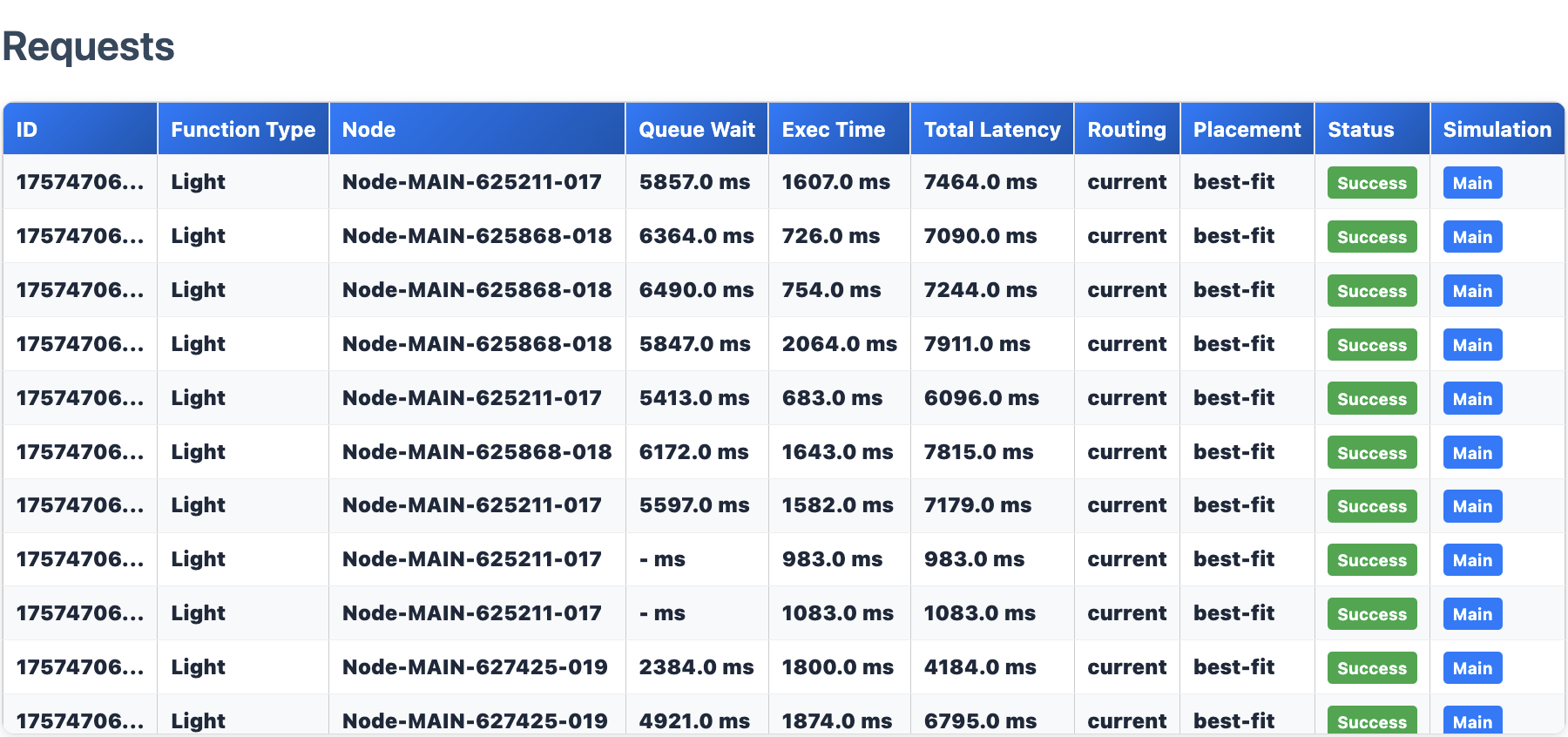}
    \caption{Live Request Metrics Collection}
    \label{fig:metrics}
\end{figure}

\subsection{Comprehensive Data Collection and Export}

The platform offers several features for data observation and export, providing multiple layers of analytical depth. Live Metrics, Fig. \ref{fig:metrics} are continuously captured and presented in detailed tables that provide a real-time snapshot of the simulation's state, including the status and resource usage of active function instances and compute nodes, as well as a history of recently completed or failed requests.

For in-depth analysis, the platform generates a variety of interactive performance charts. A dynamic bar chart, Fig. \ref{fig:latest_performance}, provides real-time insights into the average Queue Wait Time and Execution Time for requests processed within the current observation session, which can be reset for short-term experiments. A cumulative line graph continuously displays aggregate performance data over the entire simulation run, including total successful requests, total failed requests, average end-to-end latency, and average resource utilisation. A key aspect of this analysis is the cost model, which Serv-Drishti calculates for each request using a formula based on execution time and memory consumption: $(executionTimeMs / 1000) * memoryMB$, Fig. \ref{fig:cost}. This accumulated cost is tracked and provides a direct link between performance and financial metrics.

For external analysis, the platform offers the capability to export the complete cumulative dataset in standard formats such as CSV. This granular data includes timestamps for various lifecycle events, component IDs, and performance metrics, enabling researchers to perform their own statistical analysis, create custom virtualisations, and validate hypotheses. Charts can also be exported as PNG images.

\begin{figure}
    \centering
    \includegraphics[width=1\linewidth]{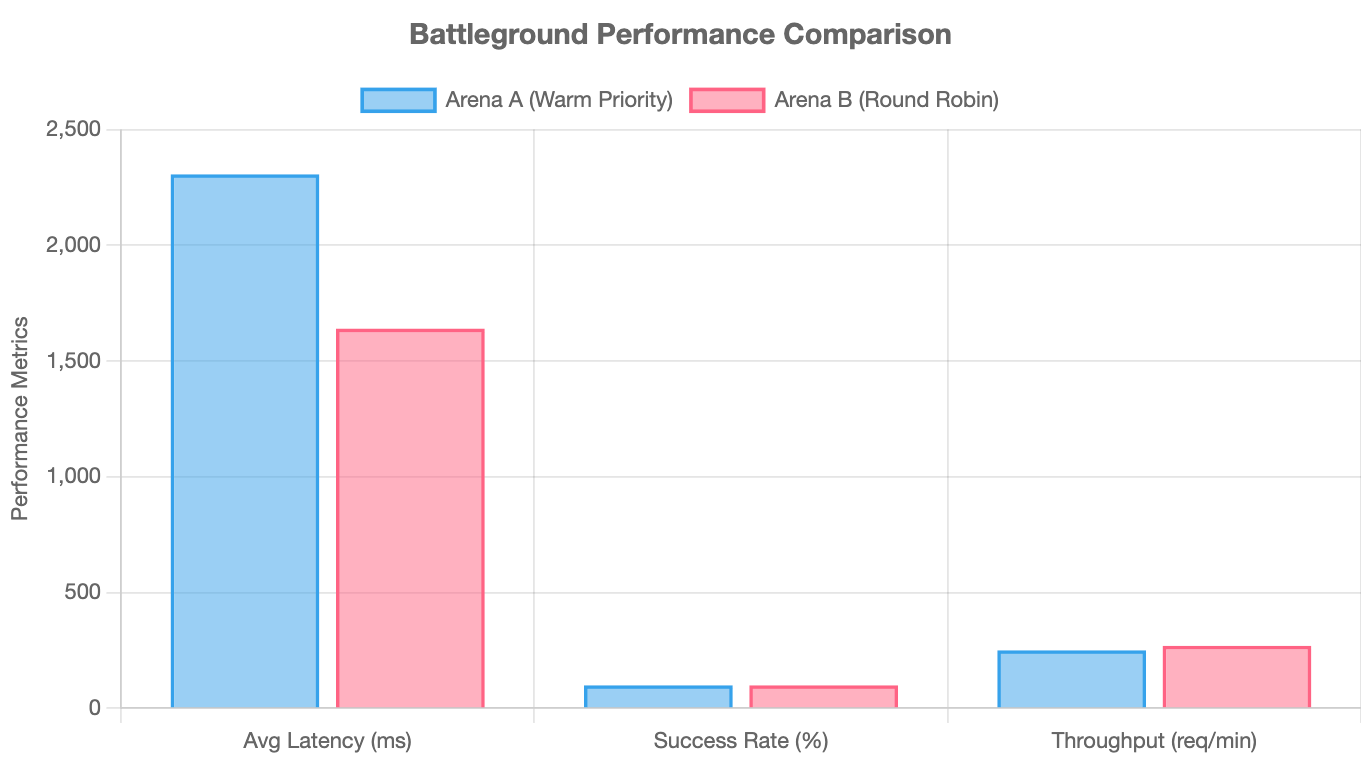}
    \caption{Battleground Performance Comparison (Routing \& Placement Strategy)}
    \label{fig:battleground}
\end{figure}

\begin{figure*}[!bpht]
    \centering
    \includegraphics[width=1\linewidth]{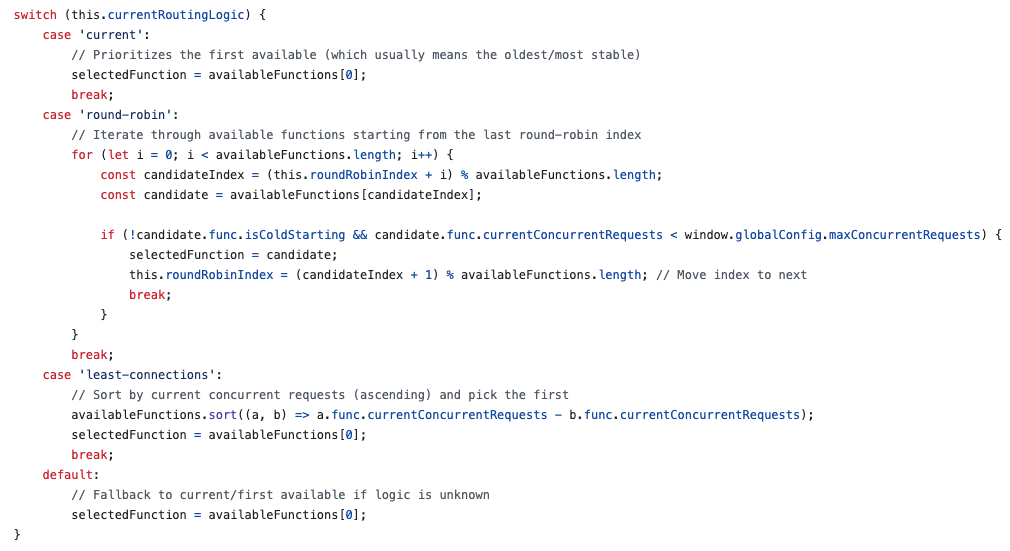}
    \caption{Serv-Drishti Request Routing Logic}
    \label{fig:routing}
\end{figure*}

\subsection{Battleground System for Comparative Analysis}

The Battleground System is another critical feature that provides a dedicated environment for comparative analysis. It runs two independent "arenas" side-by-side, each of which can be configured with a different routing or placement algorithm. Synchronised auto-requests allow for direct observation of the performance trade-offs in a single, cohesive view. The battleground generates its own set of charts for direct comparison across key metrics such as average latency, success rate, and throughput, Fig. \ref{fig:battleground}. It also features time-series charts that compare queue length, resource utilisation, active function counts, and cumulative cost between the two arenas. This powerful feature transforms Serv-Drishti from a mere demonstration tool into a versatile platform for quantitative analysis and research, providing verifiable data to back up visual observations.

\section{Extensibility and Usage}

The design of Serv-Drishti is a key consideration in its value as a research and educational platform, as its modular and open-source nature provides significant opportunities for extensibility and community contribution. It is purely implemented in JavaScript, HTML, and CSS, making the tool lightweight where it runs directly in the browser, and is highly accessible. This section outlines how users can leverage the platform's design to implement their own custom logic and details a basic guide for its practical use.

\subsection{Implementation of Custom Logic}

The clear separation between the simulation logic and the visualisation layer ensures that new features can be added without overloading the entire system and codebase. Serv-Drishti provides specific interfaces for core behaviours, allowing for the "plug-and-play" integration of custom algorithms.

To implement a new algorithm, a user can create a new function within the appropriate module (e.g., \texttt{core/placement-algorithms.js} or \texttt{core/simulation.js}), Fig. \ref{fig:routing}. This function must adhere to the established interface, taking a defined set of inputs (e.g., a list of nodes, function type) and returning a specific output (e.g., the best node for placement). The simplicity of the tech stack means no external libraries or complex build processes are needed for development.

For example, a researcher can implement a novel predictive scaling policy that provisions new function instances based on a predefined threshold of the request rate, which can be derived from the global request logs (\texttt{window.allRequestsLog}). This allows for a direct comparative analysis against the default reactive scaling policies. The core simulation logic will automatically use this new function once it is implemented at the correct position, and its performance can be visualised instantly in the platform's charts and tables.


\subsection{User Guide}

The simplicity of the UI and the browser-based implementation make Serv-Drishti highly accessible for both learning and research.
\begin{enumerate}
    \item Running a Simulation: Users can begin by simply opening the \texttt{index.html} file in a web browser, requiring no complex setup or dependencies. The user-friendly, collapsible UI panel allows for real-time adjustments of key simulation parameters, such as the \textit{auto-request rate}, \textit{cold start delay}, \textit{routing strategy} and \textit{placement strategy}. Users can also select a pre-configured demo scenario or manually trigger requests to observe the system's response in real-time. These scenarios are available from the \texttt{welcome} tab where appropriate interactive guides are also provided.
    \item Observing Results: As the simulation runs, the real-time, animated visualisations provide immediate insights into the request journey, queuing, and function state changes. For quantitative analysis, the platform’s live metrics tables and dynamic charts provide a comprehensive view of performance and resource utilisation.
    \item Data Export for Analysis: As discussed earlier, Serv-Drishti offers the capability to export all simulation data. A user can download the complete cumulative dataset in formats such as CSV, which includes granular details on each request's lifecycle events, performance metrics, function instance and compute node information. This data can then be used to perform custom statistical analysis and create visualisations beyond the tool's built-in capabilities.
\end{enumerate}

\section{Conclusions and Future Work}

The Serv-Drishti visualiser is a powerful educational and analytical tool that provides unprecedented transparency into the often-abstracted world of serverless computing. It serves as a visual and interactive guide by animating the request lifecycle, simulating complex scaling and routing logic, and demonstrating realistic failure scenarios. The platform empowers users to gain a practical and intuitive understanding of serverless platform dynamics. Its highly interactive nature and extensively configurable parameters make it suitable for various learning and experimental contexts, from a university classroom where students grasp fundamental cloud concepts to a professional architecture design session evaluating different deployment strategies. The project fills a unique and critical gap in the ecosystem of serverless tools by focusing on interactive, visual demystification for pedagogical and architectural purposes, distinguishing itself from purely programmatic simulators or real-time monitoring solutions.

Future enhancements for Serv-Drishti will build on this foundation to explore several promising directions. We plan to investigate and implement new request dispatcher strategies that leverage predictive scaling models, such as those based on machine learning, to anticipate future load and provision resources preemptively. This will allow for a direct comparative analysis against reactive scaling policies. Additionally, we will enhance the simulation model to include and visualise network latency between the API gateway, request dispatcher, compute nodes, and external services, providing a more complete picture of end-to-end latency.

To further enrich the simulation, we will expand the failure module to include more granular and complex error conditions, such as simulating specific runtime exceptions within functions and demonstrating retry mechanisms with exponential back-off or the use of dead-letter queues. We also aim to integrate a more detailed cost model to visualise the financial implications of different scaling strategies and function configurations, helping users understand how architectural decisions translate into operational costs. Finally, we will evolve the simulator to support stateful function simulation and multi-function workflows (represented as directed acyclic graphs or DAGs). This will enable the visualisation and analysis of more complex, real-world serverless workflows and their associated orchestration overheads, moving beyond the current focus on stateless functions.

\textbf{Software Availability:} The source code of Serv-Drishti simulation engine and visualiser is accessible on https://github.com/SidAg26/Serv-Drishti as an open-source tool under the Apache 2.0 license.

\bibliographystyle{plain}
\bibliography{citation}

\end{document}